\documentclass[a4,twocolumn,epj]{webofc}
\usepackage[varg]{txfonts}   
\woctitle{LHCP 2013}
\def\WmnH  {\ensuremath{\mathrm{W}(\mu\nu)\mathrm{H}}}
\def\WenH  {\ensuremath{\mathrm{W}(e\nu)\mathrm{H}}}
\def\WtnH  {\ensuremath{\mathrm{W}(\tau\nu)\mathrm{H}}}

\def\ZnnH  {\ensuremath{\mathrm{Z}(\nu\nu)\mathrm{H}}}
\def\ZllH  {\ensuremath{\mathrm{Z}(\ell\ell)\mathrm{H}}}
\def\dphiVH {\ensuremath{\Delta\phi(\mathrm{V,H})}}
\def\ptV {\ensuremath{p_{\mathrm{T}}(\mathrm{V})}}
\begin{document}
\title{Study of Higgs Production in Fermionic Decay Channels at CMS}
%
%

\author{Niklas Mohr\inst{1}\fnsep\thanks{\email{niklas.mohr@cern.ch}} 
        on behalf of the CMS Collaboration.
}

\institute{Institute for Particle Physics, ETH Z\"urich, Schafmattstrasse 20, CH-8093 Z\"urich}

\abstract{%
In these proceedings to the LHCP conference 2013 results are presented on the study of the Higgs-like particle at a mass of 125~GeV decaying into final states consisting of either $\tau^+\tau^-$, or a $b\bar{b}$ quark pair, based on the full statistics of about 24~fb$^{-1}$, collected in 2011 and 2012 at 7 and 8 TeV respectively with the CMS experiment at the Large Hadron Collider. Leptonic and hadronic decay channels for the $\tau$-lepton are included in the search. Different production channels namely gluon fusion, vector boson fusion (VBF) and associated production with W/Z bosons have been studied $\tau^+\tau^-$ final states. The $b\bar{b}$ decay channel is studied in VBF as well as in associated production with W/Z and top quarks. 
}
\maketitle
%
\section{Introduction}
\label{intro}

The recently discovered Higgs-like particle~\cite{atlasHigObs,cmsHigObs} 
has properties so far consistent with the Standard Model (SM) expectation
in terms of spin, parity and couplings~\cite{atlasCombo,cmsCombo}.
However the excess in fermionic decay channels is still not firmly established.
Therefore the study of Higgs properties in fermionic decay channels
is of great importance to confirm or disprove the SM nature of the discovered
Boson.
In the SM with the present dataset of approximately 24~fb$^{-1}$ delivered by the Large Hadron Collider
(LHC) and recorded by the CMS experiment~\cite{cmsJINST} the 
Higgs decay modes into $\tau^+\tau^-$ (denoted by $\tau\tau$ in the following) 
and $b\bar{b}$ are accessible. For a mass of 125~GeV the branching ratios are 
about 6.2\% for the $\tau\tau$ decay
and about 57\% for the $b\bar{b}$ decay.
In the $b\bar{b}$ decay channel the different production modes are very
difficult to access and suffer from very different background composition, 
therefore different analysis techniques are used to
extract the Higgs signal in each production mode. 
The studied production modes cover gluon fusion (only in $\tau\tau$), vector boson fusion (VBF) and associated production with 
W/Z bosons and top quarks (only in $b\bar{b}$). 

In these proceedings the four main analysis channels in fermionic decay modes,
$H(\tau\tau)$, $t\bar{t}H(b\bar{b})$, VBF-$H(b\bar{b})$ and $VH(b\bar{b})$
are summarized.

\section{$H(\tau\tau)$ channel}
\label{sec:Htt}

In the $\tau\tau$ channel~\cite{cmsHtt} a combined analysis on the $\tau\tau$ invariant mass
distribution is employed. The analysis uses the full 7 and 8~TeV dataset of $24.3$~fb$^{-1}$.
The combined analysis tests the following production
modes: gluon fusion, VBF and associated production with a vector boson.
Five independent $\tau$-pair final states have been studied: $\mu\tau_h$, $e\tau_h$, $e\mu$ and $\tau_h\tau_h$ channels
where $\tau_h$ denotes a reconstructed hadronic $\tau$ decay.
The one-jet category selects primarily signal events with a Higgs boson produced by gluon fusion, 
or in association with a $W$ or $Z$ boson decaying hadronically.  
Events in the VBF category are required to have two jets separated by a 
large rapidity gap, which mainly selects signal events with a Higgs boson produced by 
VBF and strongly enhances the signal contribution. 

\begin{figure}
\centering
\includegraphics[width=0.4\textwidth,clip]{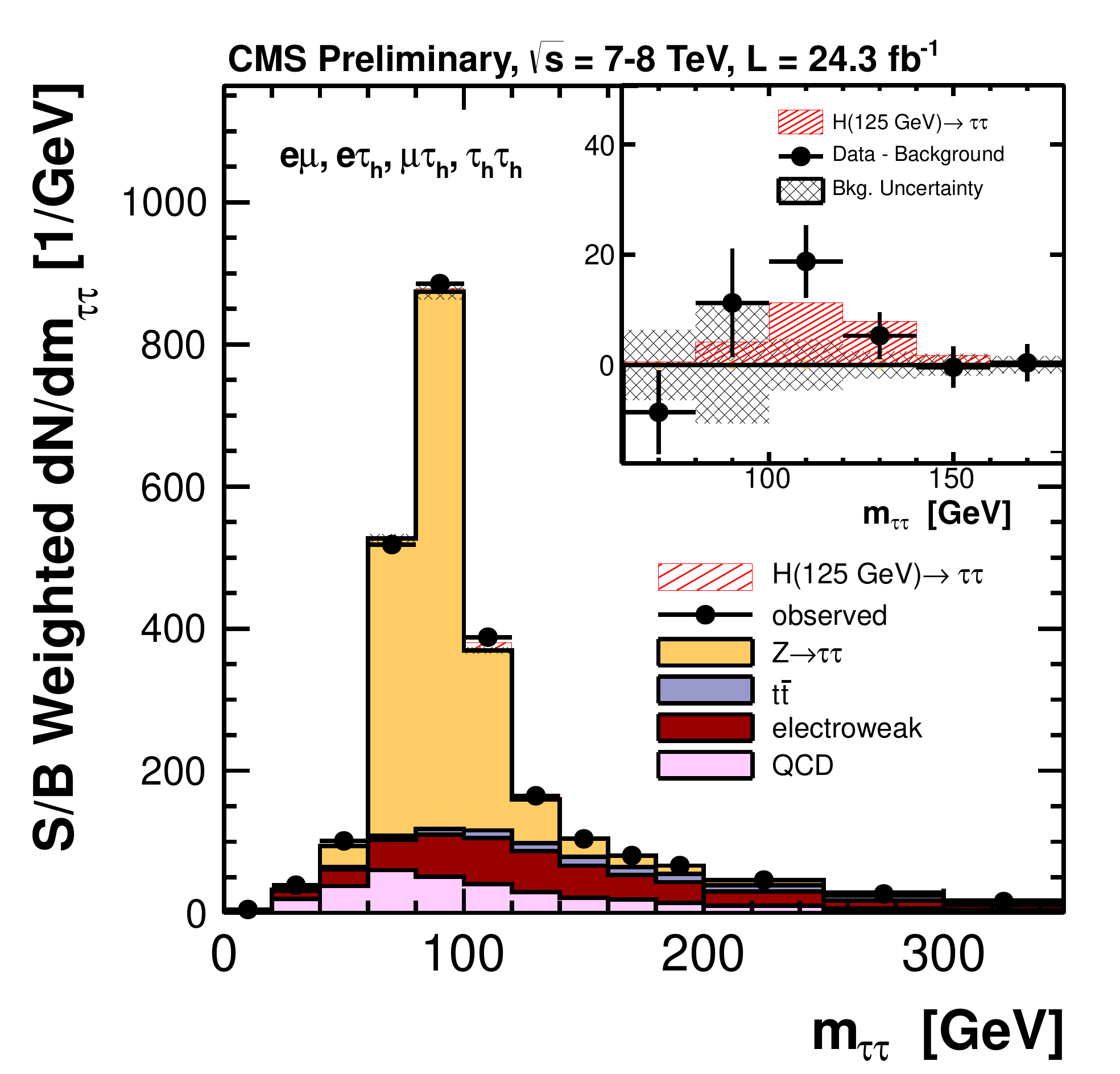}
\caption{Observed $m_{\tau\tau}$ distributions for a combination of the $\mu\tau_h$, $e\tau_h$, $e\mu$ and $\tau_h\tau_h$ channels of the $H(\tau\tau)$ analysis. The distributions obtained in each category of each channel are weighted by the ratio between the expected signal and background yields.
The insert shows the corresponding difference between the observed data and expected background distributions, together with the expected signal distribution for a standard-model Higgs signal at $m_{\rm H}=125$~GeV, with a focus on the signal region.}
\label{fig:Htt_Mass}       
\end{figure}

The event categories are defined in the following way, using jets within $|\eta|<4.7$.

\begin{itemize}
\item {\bf VBF:}
In the VBF category, two jets with $p_T>30$~GeV are required to tag the vector-boson fusion Higgs-production process.
The two jets must have an invariant mass $M_{jj}>500$~GeV and be separated in pseudorapidity by $\Delta \eta>3.5$.
A rapidity gap is defined by requiring no additional jet with $p_T>30$~GeV between the two tagging jets.
In the $e\mu$ channel, the large $t\bar{t}$ background contribution is suppressed by rejecting events containing a $b$-tagged jet of $p_T>20$~GeV.  
\item {\bf 1-jet:} Events in this category are required to have at least one jet with $p_T>30$~GeV,
not to be part of the VBF event category,
and not to contain any $b$-tagged jet with $p_T>20$~GeV.
In the $e\tau_h$ channel, the large background from  Z$\rightarrow ee$ + jets events in which an electron is misidentified as $\tau_h$ is reduced by requiring $MET>30$~GeV.
\item {\bf 0-jet:} This category contains all events with no jet with $p_T>30$~GeV, and no $b$-tagged jet with $p_T>20$~GeV. The 0-jet category is only used to constrain background normalization, identification efficiencies, and energy scales.
\end{itemize}

Backgrounds are estimated mainly from the data itself. The dominant Drell--Yan production $Z \rightarrow \tau\tau$
is estimated by an ``embedding'' technique selecting $Z \rightarrow \mu\mu$ events and replacing
the muons with simulated $\tau$ decays.
Background from $W$ + jets production is estimated
in a high-transverse mass control region
dominated by the $W$ + jets and extrapolated to the signal region
using simulated events. QCD backgrounds are estimated from same-sign events
in data. Other small background components are estimated using 
MC simulations.

The SVFit algorithm is used to improve mass reconstruction in all final states and categories
allowing a better separation between signal and background than using only the $\tau\tau$ invariant
mass from visible $\tau$ decay products by utilizing the measurement of the missing transverse energy.

Figure~\ref{fig:Htt_Mass} shows the combined observed and expected $m_{\tau\tau}$ distributions,
weighting all distributions in each category by the ratio between the expected 
signal and background yields
for this category in a $m_{\tau\tau}$ interval containing 68\% of the signal.    
It also shows the difference between the observed data and expected background distributions,
together with the expected distribution for a SM Higgs boson signal with $m_{\rm H}=125$~GeV. 

\begin{figure}
\centering
\includegraphics[width=0.4\textwidth,clip]{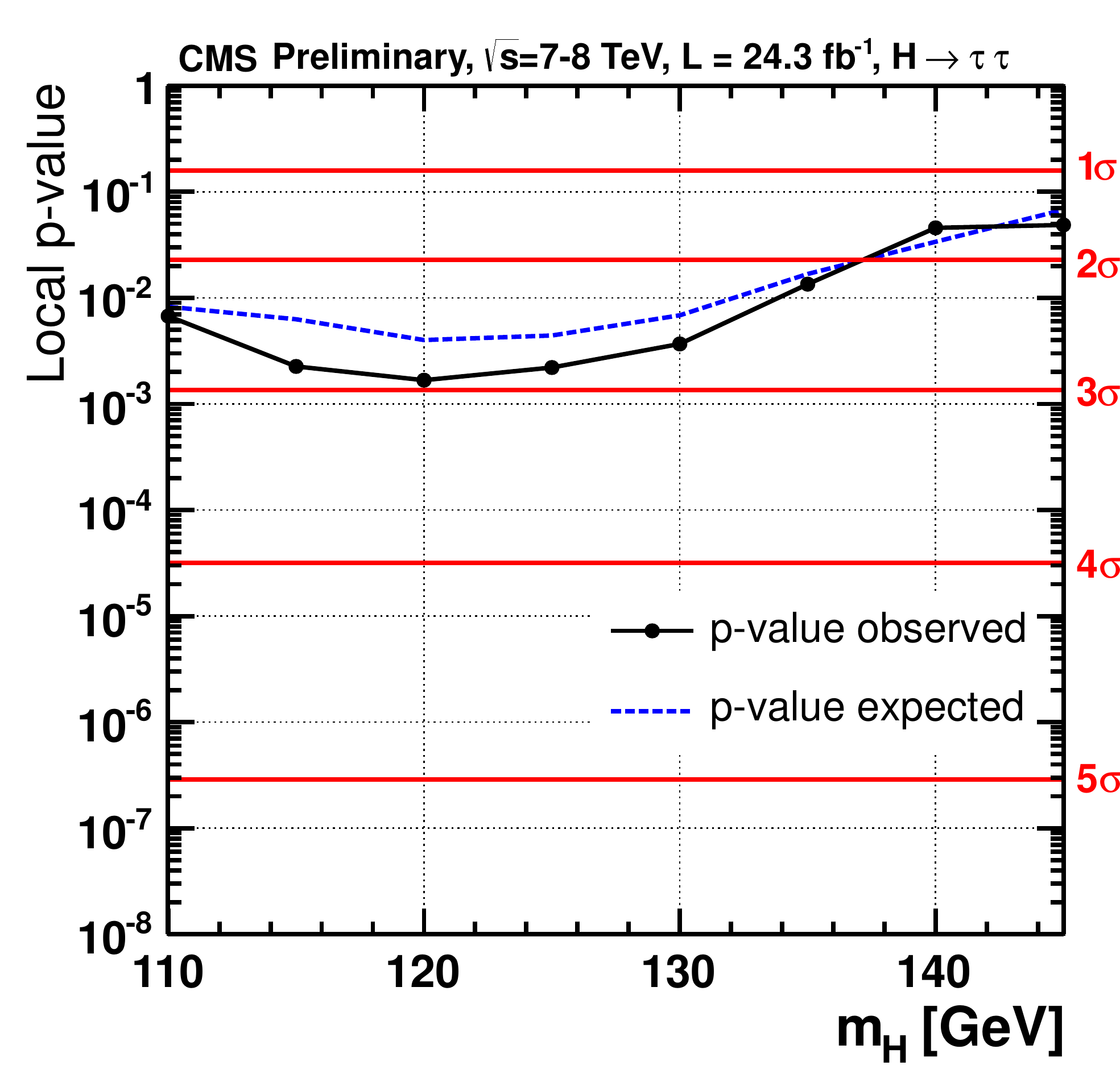}
\caption{Observed and expected p-value 1-CL$_b$, and the corresponding significance in number of standard deviations in the $H(\tau\tau)$ analysis. These results include the search for a SM Higgs boson decaying into a $\tau$ pair and produced in association with a $W$ or $Z$ boson decaying leptonically.}
\label{fig:Htt_Pvalue}       
\end{figure}

The best-fit value for the signal strength combining all channels is $\mu = 1.1 \pm 0.4$ at $m_{\mathrm H}$ = 125~GeV.
Fig.~\ref{fig:Htt_Pvalue} shows the expected and observed p-value versus Higgs boson mass. The observed excess is compatible with the presence of a standard model Higgs boson of mass $m_{\mathrm H} = 125$~GeV,
for which the local significance is $2.85\,\sigma$.

\section{$t\bar{t}H(b\bar{b})$ channel}
\label{sec:tthbb}

In the $t\bar{t}H(b\bar{b})$ channel~\cite{cmsttHbb}
the events are classified based on the decay of the associated
top quarks: lepton+jets and dilepton events. The
dominant background consists of $t\bar{t}$ events.
The analysis uses the full 7~TeV and the first $5.1$~fb$^{-1}$ of the 8~TeV dataset,
so in total an integrated luminosity of $10.1$~fb$^{-1}$.

To increase the sensitivity of the analysis
selected events are separated into different categories based on the
number of jets and $b$-tags.  For lepton+jets events, the following
seven categories are used: $\geq$6~jets + 2~b-tags, 4~jets + 3~b-tags, 5~jets +
3~b-tags, $\geq$6~jets + 3~b-tags, 4~jets + 4~b-tags, 5~jets +
$\geq$4~b-tags, and $\geq$6~jets + $\geq$4~b-tags.  For dilepton
events, only two categories are used: 2~jets + 2~b-tags and
$\geq$3~jets + $\geq$3~b-tags.

Artificial neural networks (ANNs) are used in all
categories of the analysis to further discriminate signal from
background and improve signal sensitivity. Separate ANNs are trained
for each jet-tag category, and the choice of input variables is
optimized for each as well.  The ANN input variables considered are
related to object kinematics, event shape, and the discriminant output
from the b-tagging algorithm. A total of 24 input variables has been
considered where only a subset has been used in each category.

\begin{figure}
\centering
\includegraphics[width=0.4\textwidth,clip]{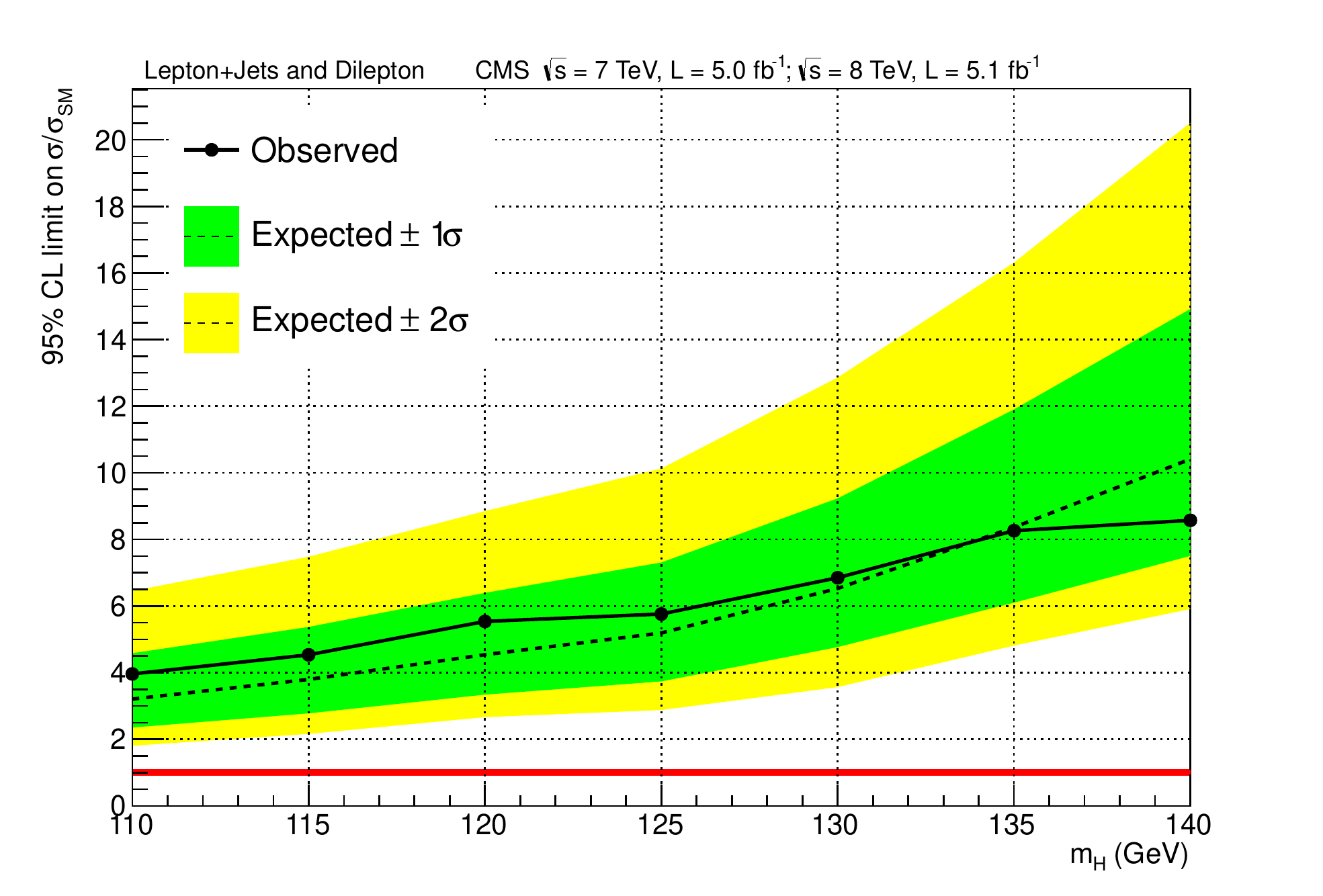}
\caption{The observed and expected 95\% CL upper limits on the signal strength parameter for lepton+jets and dilepton channels combined in the $t\bar{t}H(b\bar{b})$ channel.}
\label{fig:ttHbb_Limit}       
\end{figure}

Backgrounds are estimated from MC simulation and
the dominant systematic uncertainties originate from
the $t\bar{t}+b\bar{b}$ theoretical uncertainty
as well as from b-tagging.

The observed (expected) 95\% confidence level (CL) limit on the cross section for Higgs boson
production in association with top-quark pairs for masses from 110--140 GeV, 
have been derived using the 7~TeV and 8~TeV samples (Fig.~\ref{fig:ttHbb_Limit}).
No significant excess is found and the limit at a Higgs boson mass 
of $125$~GeV is 5.8 (5.2) times the standard model expectation.

\section{VBF-$H(b\bar{b})$ channel}
\label{sec:vbfhbb}

In the VBF-$H(b\bar{b})$ channel~\cite{cmsVBFHbb} events are characterized by
two forward quark jets and two $b$-jets originating from the decay of
the Higgs boson. The by far dominant background consists of purely QCD 
multi-jet events which already were reduced at the trigger level. 
Data at a center of mass energy of $8$~TeV corresponding to an integrated
luminosity of 19~fb$^{-1}$ have been analyzed.
The events are selected by requiring four jets with
$p_T > 85, ~70, ~60, ~40$~GeV.
The four jets are ordered in pairs labeled ``bb''
and ``qq''  alternatively with
b-tag ordering, where the ``qq'' pair is made with least b-tagged jets, and  
with $\eta$ ordering, where the ``qq'' pair is the most $\eta$-separated jet pair.
For both orderings the event selection further requires
$ m_{\rm qq}>300$~GeV and  $\Delta\eta_{\rm qq}>2.5 $.
Finally, to remove the large QCD contribution of back-to-back $bb$ pairs, events
are required to satisfy $\Delta\phi_{\rm bb}<2$, for the b-tag ordered jet pair only.

To further identify if the less b-tagged jet pair among the four leading jets
is likely to originate from the hadronization
of a light (u,d,s-type) quark, as for signal VBF tagging jets, or from gluons, as is more
probable for jets produced in QCD processes, a quark-gluon discriminator
has been applied to the b-tag sorted ``qq'' candidate jets. 

The signal extraction is performed on the $bb$ pair invariant mass distribution. 
To improve the mass resolution a neural network regression technique is employed.
The sensitivity of the analysis within pre-selection is improved by 
classifying the events based on a neural network using the characteristics of the 
``qq''  jet pair in order not to bias the $bb$ pair invariant mass distribution. 

\begin{figure}
\centering
\includegraphics[width=0.4\textwidth,clip]{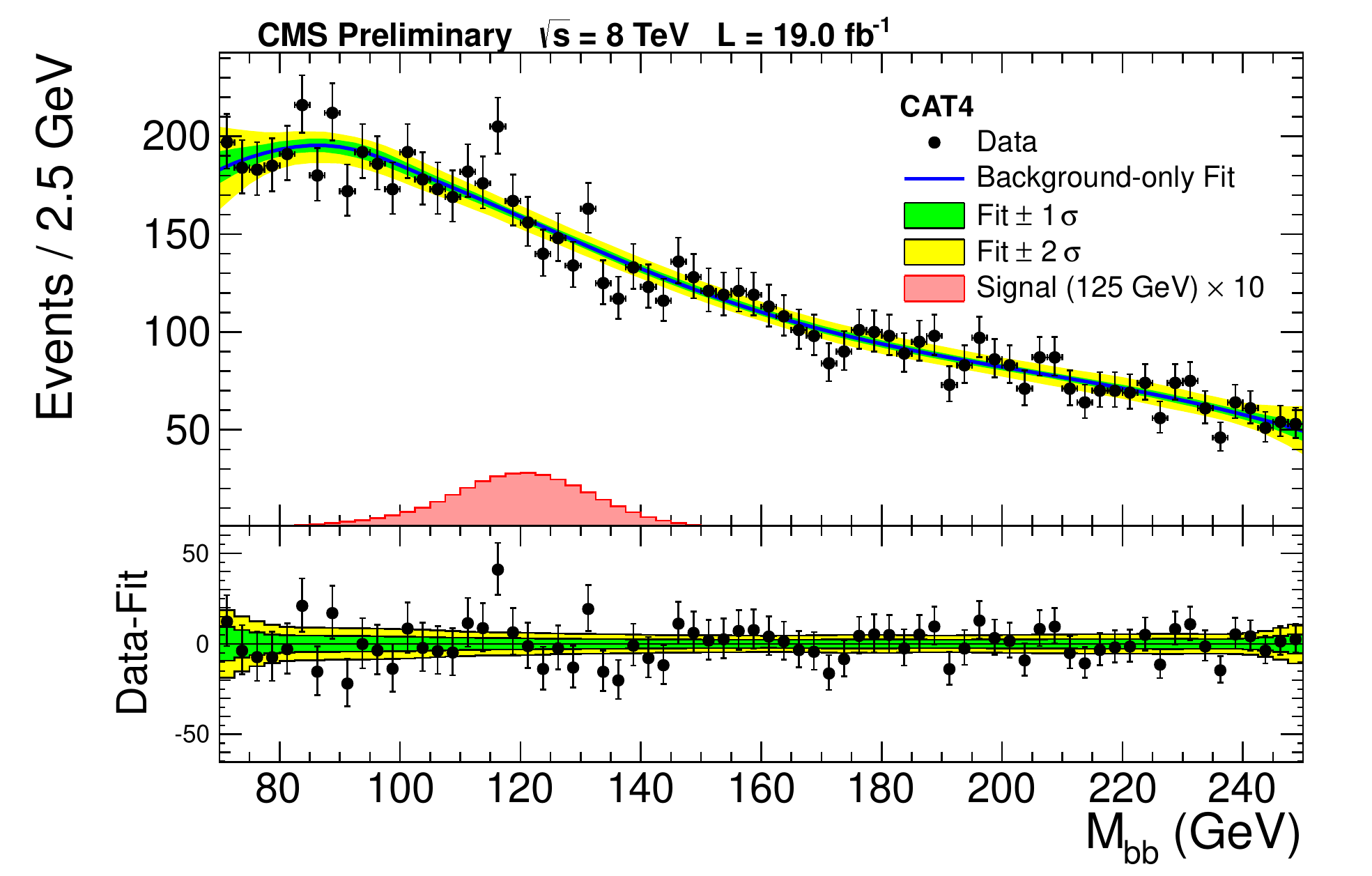}
\caption{Fit of the background model to the data $m_{bb}$ distribution in the most sensitive category of the VBF-$H(b\bar{b})$ analysis. The top panel show the fitted curve and the amplified distribution of the searched SM signal, in red.
The bottom panel shows the data minus fit residuals, together with the fit one and two sigma uncertainty band.}
\label{fig:VBF_Mass}       
\end{figure}

Data are categorized in four categories and a polynomial
of fifth degree is used to model the QCD continuum background. A fit of the 
$bb$ pair invariant mass distribution in the most 
sensitive category is shown in Fig.~\ref{fig:VBF_Mass}.

To validate the search strategy the $m_{bb}$ fit is performed in the same way as for the
Higgs boson search, but trying to extract VBF Z production with $Z\rightarrow b\bar{b}$. The data
yields 2844$\pm$1127 events in the Z peak, i.e. a fit of the Z peak with a significance of 2.5
standard deviations, in agreement with expectations derived using Monte~Carlo pseudo experiments.

\begin{figure}
\centering
\includegraphics[width=0.4\textwidth,clip]{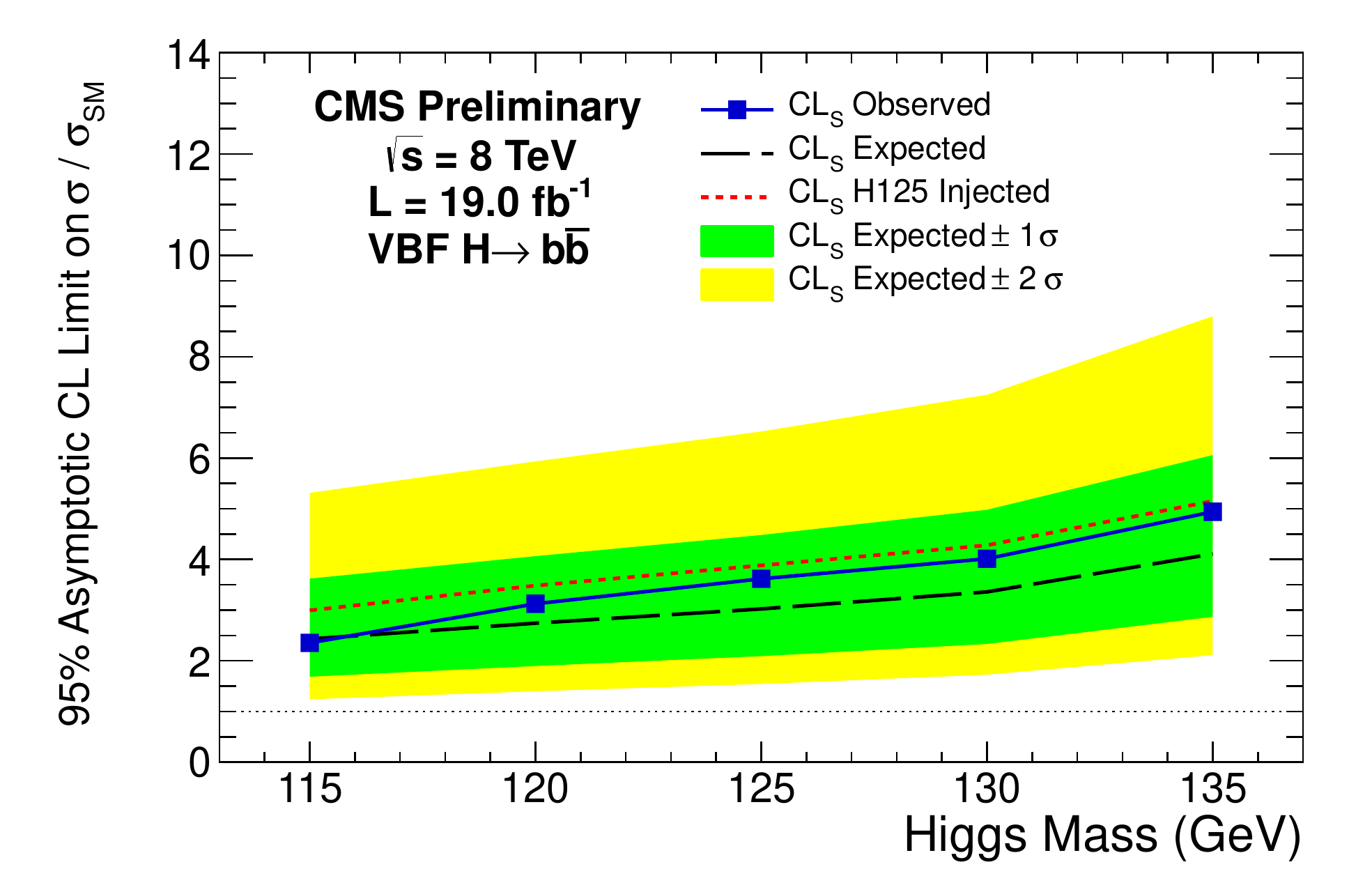}
\caption{Expected and observed 95\% confidence level limits on the signal cross section in units of the SM
expected cross section for the VBF-$H(b\bar{b})$ analysis, as a function of the Higgs boson mass, 
including all four higher ANN event categories.
The limits expected in the presence of a SM Higgs boson with mass 125~GeV are indicated by the dotted curve.}
\label{fig:VBF_Limit}       
\end{figure}

Based on all four categories upper limits, at the 95\% confidence level, on the production cross section times the 
branching ratio, with respect to the expectations for a standard model Higgs boson,
are derived for a Higgs boson in the mass range 115--135 GeV.
In this range, the expected upper limits in the absence of a signal vary 
from 2.4 to 4.1 times the standard model prediction,
while the corresponding observed upper limits vary from 2.4 to 5.2 (Fig.~\ref{fig:VBF_Limit}).

At a Higgs boson mass of 125 GeV the expected limit is 3.0 and the observed limit is 3.6.
The fitted signal strength is $\mu=\sigma/\sigma_{\rm SM}=0.7\pm1.4$.

\section{$VH(b\bar{b})$ channel}
\label{sec:vhbb}

In the $VH(b\bar{b})$ channel~\cite{cmsVHbb} the full LHC dataset of
$24$~fb$^{-1}$ at 7~TeV and 8~TeV has been analyzed. The following
modes are included in the search:
$\mathrm{W}(\mu\nu)\mathrm{H}$, $\mathrm{W}(e\nu)\mathrm{H}$,
$\mathrm{W}(\tau\nu)\mathrm{H}$, $\mathrm{Z}(\mu\mu)\mathrm{H}$,
$\mathrm{Z}(ee)\mathrm{H}$ and $\mathrm{Z}(\nu\nu)\mathrm{H}$, all
with the Higgs boson decaying to $b\bar{b}$.

The background processes to VH production originate from vector-boson+jets (V+jets), $t\bar{t}$,
single-top and dibosons (VV) production. Except for diboson production, these processes have
production cross sections that are several orders of magnitude larger
than Higgs boson production. The diboson production cross section is
only a few times larger than the production cross
section for VH and, given the nearly identical final state for VZ with
$\mathrm {Z}\rightarrow b\bar{b}$, this process provides a
benchmark against which the Higgs boson search strategy is tested.

\begin{figure}
\centering
\includegraphics[width=0.4\textwidth,clip]{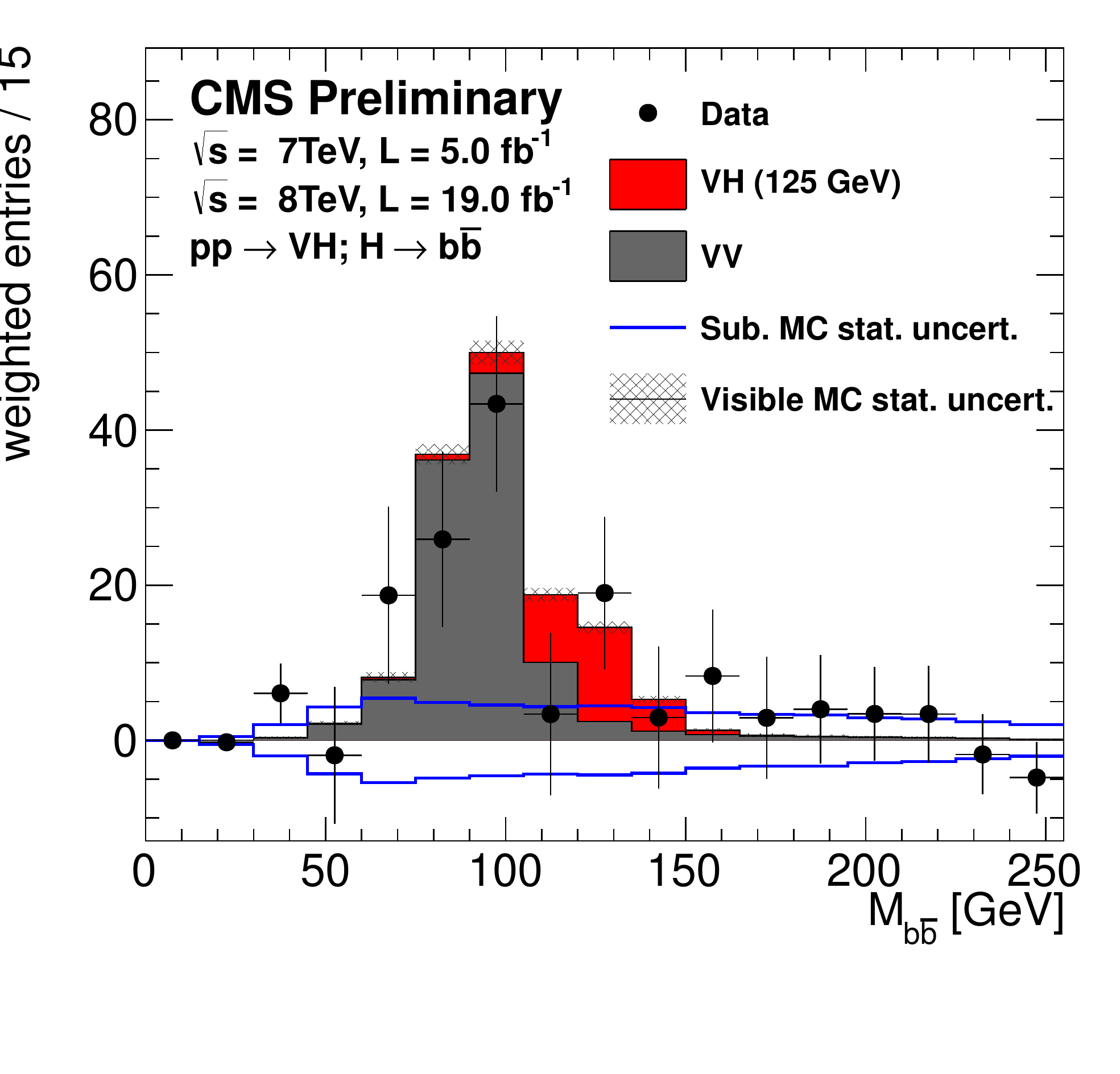}
\caption{Weighted dijet invariant mass distribution, combined for all
channels of the $VH(b\bar{b})$ analysis. For each channel, the relative weight
of each \ptV\ bin is obtained from the ratio of the expected number of
signal events to the sum of expected signal and background events in a
window of $m_{b\bar{b}}$ values between 105 and 150~GeV. The expected signal
used corresponds to the production of a Higgs boson with a mass of 125~GeV. The weight for the
highest \ptV\ bin is set to 1.0 and all other weights are adjusted proportionally.
All backgrounds, except
dibosons, have been subtracted. The solid histograms for the
background and the signal are summed cumulatively. The data is represented by points
with error bars.}
\label{fig:VHbb_MJJ}       
\end{figure}

Backgrounds are reduced by
requiring a boost of the $p_T$ of the vector boson, \ptV.
In that case the Higgs and Vector boson recoil away from each other with a 
large azimuthal opening angle,  \dphiVH,
between them.  For each mode, different
regions of \ptV\ boost are considered.
Due to different signal and background
composition, each boost region has different sensitivity and
the analysis is performed separately in each region. The
results from all regions are then combined for each channel. The
``low'', ``intermediate'', and ``high'' boost regions
for the \WmnH\ and \WenH\ channels are $100<\ptV<130$~GeV,  $130<\ptV<180$~GeV, and
$\ptV>180$~GeV.
For the \WtnH\ a single $\ptV>120$~GeV region is considered.
For the \ZnnH\ channel the ``low'', ``intermediate'', and ``high'' boost
regions are  $100<\ptV<130$~GeV, $130<\ptV<170$~GeV and
$\ptV>170$~GeV, and for the \ZllH\ channels, the ``low'' and ``high''
regions are $50<\ptV<100$~GeV
and $\ptV>100$~GeV.

The Higgs boson mass resolution
is improved by applying a bosted decision tree (BDT) regression technique. Using this a
further correction, beyond the standard CMS jet energy corrections, for
individual $b$-jets improves the $b\bar{b}$ invariant mass resolution by around 15\%.

To estimate the backgrounds a set of simultaneous fits is
performed to several distributions of discriminating
variables in the control regions, separately in
each channel, to obtain consistent scale factors by which the Monte
Carlo yields are adjusted.

In each of the \ptV\ categories a boosted decision tree discriminator is used
to distinguish signal from background events. Input variables
consist of topological information, b-tag related quantities and kinematic quantities.
The most important variable in the analysis is the mass
of the $b\bar{b}$ system (see Fig.~\ref{fig:VHbb_MJJ}), which is also used as input to the BDT.
To extract the VH signal a combined fit to all BDT
distributions is employed.

As a validation of the multi-variate approach to this analysis, these
BDT discriminants are also trained to find diboson signals (ZZ and WZ, with
Z$\to b\bar{b}$) rather than the VH
production signal. Using this Diboson BDT the VZ process
with respect to the NLO VZ cross-section,
is measured to be  ${1.19}_{-023}^{+0.28}$
with a significance exceeding $7\sigma$. 

Additionally a shape analysis on the mass 
as a single variable is used to perform a cross-check analysis
of the BDT for both VZ and VH signals, yielding consistent results
for both cases.

\begin{figure}
\centering
\includegraphics[width=0.4\textwidth,clip]{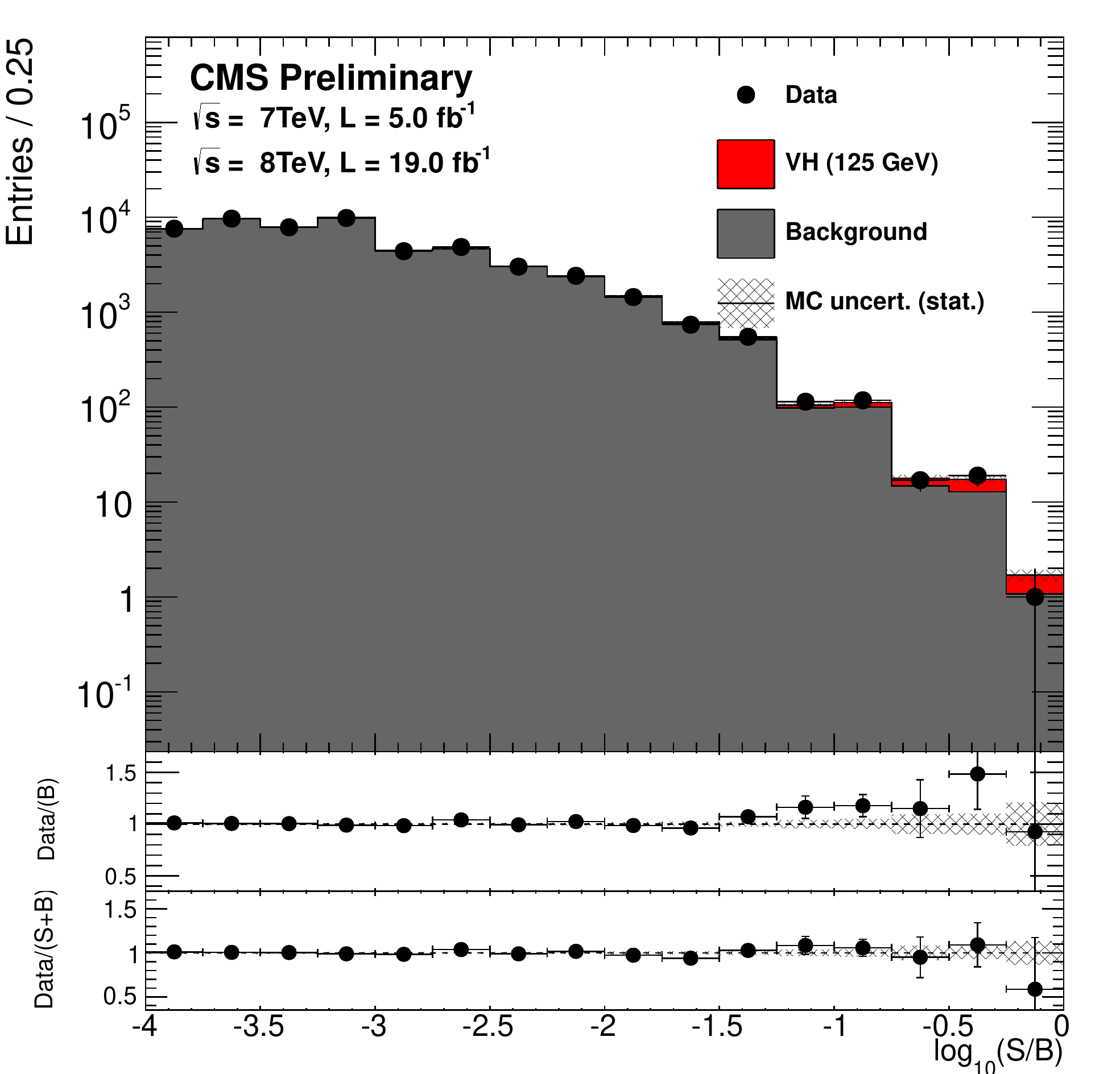}
\caption{Combination of all BDT discriminants of the $VH(b\bar{b})$ analysis into a single distribution where all
events, for all channels, are sorted in bins of similar expected signal-to-background
ratio, as given by the value of the output of their corresponding BDT
discriminant (trained with a Higgs boson mass of 125~GeV).
The two bottom insets show the ratio of the data to the background-only
prediction (above) and to the predicted sum of background plus signal
(below).}
\label{fig:VHbb_BDT}       
\end{figure}

A summary of the BDT analysis is shown in Fig.~\ref{fig:VHbb_BDT}
where all bins in the different categories are shown sorted 
by their $s/b$. An excess is visible in the most significant
bins of the analysis.

\begin{figure}
\centering
\includegraphics[width=0.4\textwidth,clip]{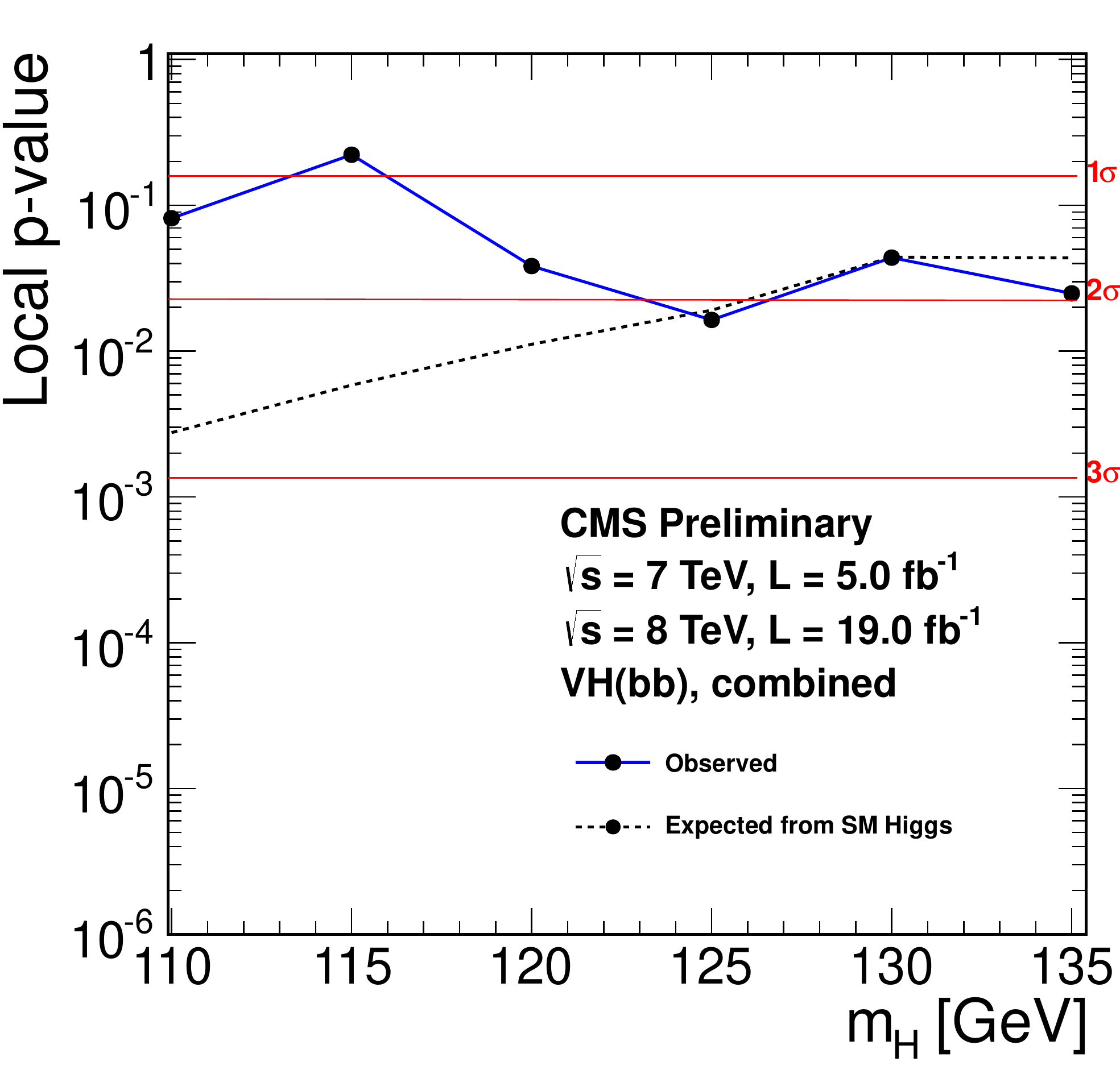}
\caption{p-values for background fluctuations to account for
the observed excess of events in the data in the $VH(b\bar{b})$ analysis.}
\label{fig:VHbb_Pvalue}       
\end{figure}

Upper limits, at the 95\% confidence level, on the
   VH production cross section times the $\mathrm{H} \to b\bar{b}$
   branching ratio, with respect to the expectations for a standard
   model Higgs boson, are derived for a Higgs boson in the mass range
   110--135~GeV.
   In this range, the observed upper limits vary from 1.1 to 3.1 times the SM prediction; the
   corresponding expected limits vary from 0.7 to 1.5. At a Higgs
   boson mass of 125~GeV the observed limit is 1.89 while the expected
   limit is 0.95. An excess of events is observed above the expected
   background with a local significance of 2.1 standard deviations,
   which is consistent with the expectation from the production of the
   SM Higgs boson (Fig.~\ref{fig:VHbb_Pvalue}). The signal strength corresponding to this excess, relative to that of
the SM Higgs boson, is  ${1.0}\pm{0.5}$.

\section{Summary}

Four analysis of Higgs boson production in fermionic decay channels have been presented.
Different production channels namely gluon fusion, vector boson fusion (VBF) and associated production with W/Z bosons have been studied in the $\tau\tau$ mode. The $b\bar{b}$ decay channel has been studied in VBF as well as 
in associated production with W/Z and top quarks.
All derived limits, significances of the excesses and signal strength with respect to the SM expectations
in the four analysis are summarized in Tab.~\ref{tab:allChan}.

\begin{table}[h]
\centering
\caption{Observed 95\% CL limits (expected in parentheses), significance and signal strength ($\mu$-value) with respect to the SM expectation in the four analysis channels.}
\label{tab:allChan}      
\begin{tabular}{lccc}
\hline
Channel & Limit & Significance & $\mu$-value  \\\hline
$H(\tau\tau)$ & 2 (0.83) & 2.9 & $1.1\pm0.4$ \\
$VH(b\bar{b})$ & 1.9 (0.95) & 2.1 & $1.0\pm0.5$ \\
VBF-$H(b\bar{b})$ & 3.6 (3.0) & - & $0.7\pm1.4$ \\
$t\bar{t}H(b\bar{b})$ & 5.8 (5.2) & - & - \\ \hline
\end{tabular}
\end{table}

In the $H(\tau\tau)$ and $VH(b\bar{b})$ first hints of a signal are emerging
at a significance above $2~\sigma$. Individually non of the channels is able 
to claim an evidence for Higgs decays into fermions
at present.

However, a combination of the $H(\tau\tau)$ and the $VH(b\bar{b})$ analysis yields an observed significance
of 3.4~$\sigma$~\cite{cmsCombo} at a Higgs boson mass of 125~GeV, which can be interpreted as 
evidence for decays of the Higgs boson into fermions.

Nevertheless a larger dataset will be needed to firmly
establish all Higgs boson decay and production processes in fermionic decay channels. This dataset
is expected to be collected at a higher center of mass energy during the LHC Run~2 starting in
2015.
%
%
%

\end{document}